\documentclass{jfm}
\usepackage{graphicx}
\usepackage{epstopdf, epsfig}
\usepackage[amssymb]{SIunits}
\usepackage{color}
\usepackage{amsmath}
\usepackage[normalem]{ulem}
\usepackage{pdflscape}

\newcommand{\makeblue}[1]{\textcolor{blue}{#1}}

\shorttitle{Thermal spectra in turbulent bubbly flows}
\title{The emergence of bubble-induced scaling in thermal spectra in turbulence}

\author{On-Yu Dung\aff{1},
  Pim Waasdorp\aff{1},
  Chao Sun\aff{1,3},
 Detlef Lohse\aff{1,2},
 \and Sander G. Huisman\aff{1}\corresp{\email{s.g.huisman@utwente.nl}}}

\affiliation{\aff{1}Physics of Fluids Group, J. M. Burgers Center for Fluid Dynamics, and Max Planck Center Twente, Faculty of Science and Technology, University of Twente, P.O. Box 217, 7500 AE Enschede, The Netherlands
\aff{2}Max Planck Institute for Dynamics and Self-Organization, Am Fassberg 17, 37077 G\"ottingen, Germany
\aff{3}Center for Combustion Energy, Key Laboratory for Thermal Science and Power Engineering of Ministry of Education, Department of Energy and Power Engineering, Tsinghua University, 100084 Beijing, China}

\begin{document}

\maketitle

\begin{abstract}
We report on the modification of the spectrum of a passive scalar inside a turbulent flow by the injection of large bubbles. While the spectral modification through bubbles is well known and well analyzed for the velocity fluctuations, little is known on how bubbles change the fluctuations of an approximately passive scalar, in our case temperature. Here we uncover the thermal spectral scaling behavior of a turbulent multiphase thermal mixing layer. The development of a $-3$ spectral scaling is triggered. By injecting large bubbles ($\text{Re}_{\text{bub}} = \mathcal{O}(10^2)$) with gas volume fractions $\alpha$ up to 5\%. For these bubbly flows, the $-5/3$ scaling is still observed at intermediate frequencies for low $\alpha$ but becomes less pronounced when $\alpha$ further increases and it is followed by a steeper $-3$ slope for larger frequencies. This $-3$ scaling range extends with increasing gas volume fraction. The $-3$ scaling exponent coincides with the typical energy spectral scaling for the velocity fluctuations in high Reynolds number bubbly flows. We identify the frequency scale of the transition from the $-5/3$ scaling to the $-3$ scaling and show how it depends on the gas volume fraction.
\end{abstract}


\section{Introduction}
\label{sec:intro}

In fully developed turbulent flows, universality is observed for velocity and temperature fluctuations for the second order structure functions and the corresponding spectra. According to the Kolmogorov--Obukhov--Corrsin theory (\cite{Kol41,Obuk49,Corrsin51}) and verified extensively in experiment and numerical simulations, the universal scaling of the spectrum of a scalar exhibits a $-5/3$ scaling ($E_{\theta}(k) \propto k^{-5/3}$) in the inertial-convective range for high enough Reynolds and P\'eclet number (see e.g. \cite{Monin75}). Here $k$ is the wavenumber and $\theta$ is a passive scalar (\cite{Monin75}) such as the concentration of a chemical substance or the temperature, in case that buoyancy is not relevant. If additional complexity is added to the system, different scaling phenomena can appear, such as having an active scalar introducing buoyancy effects (e.g. Rayleigh-B\'enard convection) (\cite{lohse2010small}), adding visco-elasticity to the carrier liquid (\cite{Steinberg2021AnnuRev}), or by adding finite particles (\cite{RissoAnnuRev2018} and \cite{BrandtColettiARFM2022}) or bubbles (\cite{Balachandar:2010AnnuRev,RissoAnnuRev2018,LohsePRF2018,MathaiAnnuRev2020}). In this work, we do the latter and introduce a large number of finite-size rising bubbles to the liquid flow (\cite{Balachandar:2010AnnuRev,RissoAnnuRev2018,LohsePRF2018,MathaiAnnuRev2020}). Large rising bubbles are defined here as having a bubble Reynolds number $\text{Re}_{\text{bub}}=V_r d /\nu = \mathcal{O}(10^2)$, where $V_r$ is the bubble rise velocity relative to the mean vertical velocity of the carrier liquid, $d$ is the mean area-equivalent bubble diameter (as compared to a spherical bubble), and $\nu$ is the kinematic viscosity of the carrier liquid. 

First observed and theoretically addressed in \cite{Lance:2006ht}, the introduction of bubbles lead to the emergence of a $k^{-3}$ scaling of the velocity fluctuation (energy) spectra $E_{u}(k)$. They argued that in a statistical steady state, the $k^{-3}$ scaling in the spectrum can be achieved by balancing the viscous dissipation and the energy production due to the rising bubbles in the spectral space (\cite{Lance:2006ht}). The $k^{-3}$ scaling can also result from spatial and temporal velocity fluctuations in homogeneous bubbly flows (\cite{RissoAnnuRev2018}), and is found to be robust and is observed for a homogeneous bubble swarm for a wide range of parameters: $10 \leq \text{Re}_{\text{bub}} \leq 1000$  and $1 \leq \text{We} \leq 4$ where $\text{We}$ is the Weber number (\cite{RissoAnnuRev2018,PandeyJFM2020}). Recently, \cite{Ma2022JFM} employed particle shadow velocimetry in bubble-laden turbulent flow and obtained a high-resolution two-dimensional velocity field. They revealed longitudinal and traverse velocity structure functions up to twelfth order, in which the second order statistics (one-dimensional longitudinal energy spectra) also show a $-3$ scaling. In contrast, point particle simulations do not show $k^{-3}$ scaling (\cite{Mazzitelli2009hn}), because the wakes behind the bubbles which are obviously absent for point particles are crucial for the emergence of a $-3$ scaling (in either frequency or wavenumber space). 

While there are abundant studies on the energy spectra in high-$\text{Re}$ bubbly flow, the study of scalar spectra in such flow is limited. \cite{Almeras2016TimeResolved} investigated the time-resolved concentration fluctuations of a fluorescent dye (passive scalar) in a confined bubbly thin cell in which the scalar spectral scaling $f^{-3}$ is observed, where $f$ is the frequency. For the three-dimensional case, \cite{Gvozdic:2018JFM} investigated the scalar spectra in a bubbly column in a vertical convection setup, but they were unable to resolve the frequencies related to the $-3$ scaling of the energy spectra in bubble-induced turbulence (\cite{Gvozdic:2018JFM}). The obvious downside of using dyes is that the water gets contaminated and only very short runs can be performed before the water has to be completely replaced. By using temperature as the approximately passive scalar, and the active cooling of the setup during the return stage, it is possible to maintain an overall constant temperature of the setup, allowing for continuous measurements for hours or days if needed.

In the present experimental work, by utilizing a fast-response temperature probe, we are able to capture the gradual change of the \textit{passive scalar (thermal)} spectral scaling for increasing bubble concentration $\alpha$ in a bubbly turbulent thermal mixing layer. In addition, we will investigate the transition scale of the $-3$ scaling as it illuminates the physical origin of the $-3$ subrange of both the scalar and the energy spectra. The transition frequency to the $-3$ subrange for energy spectra indeed has yet to be exactly identified (\cite{RissoAnnuRev2018}).

\section{Experimental setup and methods}\label{sec:ch3ExpSetupAndMethod}

\subsection{Experimental facility}\label{sec:Expsetup}

\begin{figure}
  \centerline{\includegraphics[width=1\columnwidth]{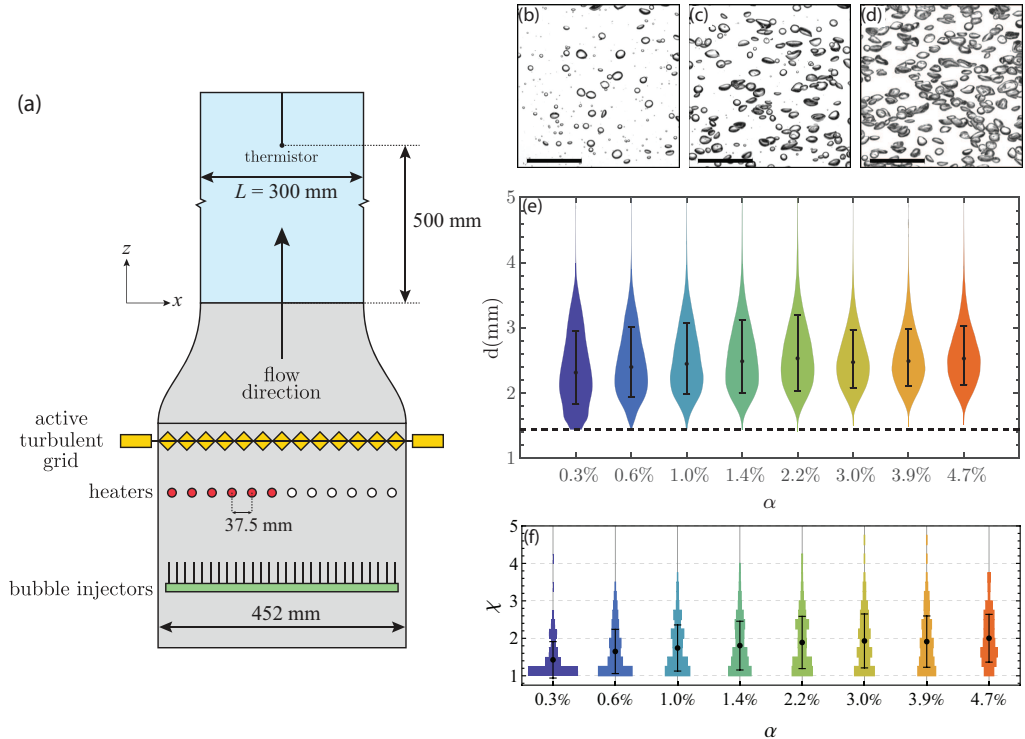}}
  \caption{(a) Schematic of the experimental setup used in the current study. Bubbles are injected by 120 needles (green). The setup is equipped with 12 heaters. A power of \unit{2250}{\watt} is supplied to the left half of the heaters (red), while the right half (white) is left unpowered, creating a turbulent thermal mixing layer. Each heating cartridge (Watlow, Firerod \texttt{J5F-15004}) has a diameter of \unit{12.7}{\milli\meter}. The center-to-center distance between two consecutive heaters is \unit{37.5}{\milli\meter}. The turbulence is stirred by an active grid with a total of 15 engines connected to diamond flaps (yellow) (\cite{Gvozdic:2019RSI}). The measurement section (blue area) has a height of \unit{1.00}{\meter}, a width of \unit{0.3}{\meter}, and is \unit{0.04}{\meter} thick. (b,c,d) High speed snapshots of the bubble flow, as seen from the side for the volume fractions $\alpha= 0.3\%, 1.4\%$, and $4.7\%$. Scale bars of \unit{20}{\milli\meter} are added to each figure. (e) Distribution charts of the bubble diameter obtained using image analysis of high-speed footage for a variety of gas volume fractions $\alpha$. The resolution of the footage prohibits us to measure the size of bubbles (dashed line)  $<\unit{1.4}{\milli\meter}$. Dots indicate the mean bubble diameter $d$. (f) Histograms of the bubble aspect ratio $\chi$ obtained by fitting the bubbles on the high-speed footage with ellipses. $\chi$ is defined as the ratio of the semi-major axis divided by the semi-minor axis. Each black dot represents the mean aspect ratio $\chi$ and each error bar represent one standard deviation of the distribution. See also Table \ref{tab:bubbleProperties_fitResults} for the details of the bubble properties.}
\label{fig:setup} 
\end{figure}

We utilize the Twente Mass and Heat Transfer Tunnel (\cite{Gvozdic:2019RSI}) which creates a turbulent vertical channel flow, and allows for injecting bubbles and heating the liquid in a controlled manner. The turbulence is actively stirred using an active grid, see Fig.~\ref{fig:setup}a. Using this facility we create a turbulent thermal mixing layer (\cite{Kops:PoF2000}) in water with a mean flow velocity of approximately \unit{0.5}{\meter\per\second}, resulting in a Reynolds number of $\text{Re} = 2 \times 10^4$. The global gas volume fraction $\alpha$ in the measurement section is measured by a differential pressure transducer of which the two ends are connected to the top and bottom of the measurement section. A fast-response thermistor (Amphenol Advanced Sensors type \texttt{FP07}, with a response time of \unit{7}{\milli\second} in water) is placed into the middle of the measurement section in order to measure the temperature. An AC bridge with a sinusoidal frequency of \unit{1.3}{\kilo\hertz} drives the thermistor in order to reduce electric noise. The bridge potential is measured by a lock-in amplifier (Zurich Instruments \texttt{MFLI}) with the acquisition frequency of \unit{13.4}{\kilo\hertz} and such that the cut-off frequency that is used is $\unit{(13.4/2)}{\kilo\hertz}$, which is far away from the effective range of the thermistor which is limited by the thermal capacity of the probe. A temperature-controlled refrigerated circulator with water bath (PolyScience \texttt{PD15R-30}) with a temperature stability of \unit{\pm 0.005}{\kelvin} is used for calibrating the thermistor. We employ the temperature-resistance characteristic equation proposed by \cite{Steinhart1968} over a short range of temperature of interest (in which the terms are kept only up to the first order to avoid over-fitting) in order to convert the measured resistance of the thermistor to the corresponding temperature. The thermistor has a relative precision of \unit{1}{\milli\kelvin}. With relative we mean here relative to a baseline temperature, i.e.\ a temperature difference ($\Delta T$), as we are mainly interested in temperature \textit{changes}, which have a different precision than $T$ itself. The typical large-scale temperature difference from the heated side to the non-heated side at the mid-height of the measurement section of height $h =  \unit{1}{\meter}$ is about \unit{0.2}{\kelvin}. The working liquid is decalcified tap water at around \unit{22.7}{\celsius}, with a Prandtl number $\text{Pr}=\nu/\kappa=6.5$, where $\nu$ and $\kappa$ are the kinematic viscosity and thermal diffusivity of water, respectively.



For the velocity measurements, we employ Laser Doppler Anemometry (LDA) (Dantec) and constant temperature anemometry (CTA) using a hot film sensor (Dantec \texttt{55R11}, \unit{70}{\micro\meter} diameter with a \unit{2}{\micro\meter} nickel coating, \unit{1.3}{\milli\meter} long and up to \unit{30}{\kilo\hertz} frequency response) placed in the middle of the tunnel. CTA is used for calculating the velocity spectra and the corresponding method employed in this work can be found in Appendix \ref{sec:SpectralMethod}. We note that since a CTA measurement requires negligible temperature variation of the flow, the velocity is measured when the heaters are switched off, but left in place such as to not alter the incoming flow. The velocity statistics are not influenced by the temperature because the fluctuations in the flow are completely dominated by the turbulence generated by the active grid and the upward mean flow rather than by buoyancy caused by density differences due to the temperature differences, i.e.,\ it is an approximately passive scalar. To substantiate this statement, we must compare the relative importance of the buoyancy force (parallel to the streamwise direction) to the inertial force due to the advection of a passive scalar. The ratio of these forces equals the ratio of the Grashof number $\text{Gr}$ and the square of the Reynolds number $\text{Re}_{0.5h}^2$ (\cite{BLtheory9th}). Here we base these numbers on the vertical distance between $\pm 0.25h$ from the middle of the measurement section, the temperature difference across such height $\Delta_{0.5h} \leq \unit{50}{\milli\kelvin}$ and the mean liquid velocity $U = \unit{0.5}{\meter {\second}^{-1}}$. This gives $\text{Gr}/\text{Re}_{0.5h}^2 =0.5hg\beta\Delta_{0.5h}/U^2 \leq \mathcal{O}(10^{-3})$, where $g$ and $\beta$ are the gravitational constant and volumetric expansion coefficient of water, respectively. The small value of this ratio implies that the temperature can be seen as passive scalar under our flow conditions.

\subsection{Measurement conditions: single-phase turbulent characterisation}\label{sec:MeasureCondition}

Next, we present the turbulent characteristics in the single-phase turbulent thermal mixing layer. Firstly, we characterise the velocity field. Laser Doppler anemometry (LDA) (Dantec) is used for measuring the horizontal ($u_x$) and vertical ($u_z$) velocity in order to characterise the velocity fluctuations. Throughout the manuscript we will use a prime ($'$) to denote the standard deviation of any quantity, and therefore the standard deviation of the velocity in the single-phase case is $u_0'$ and we approximate $u_0'\approx u_z'$. In the single-phase case, at the middle of the tunnel, $u_x'/u_z' = 0.69 \pm 0.07$, where the statistical uncertainty is one standard deviation. This is not ideally isotropic. Therefore, for single-phase case, the estimation of the relevant scales using the assumption of isotropy below is only a rough approximation. The viscous dissipation rate $\epsilon$ is estimated by finding the longitudinal second order structure function $D_{LL}(r)$ in single-phase flow, where $r$ is the spatial distance between two points (measured by CTA and used Taylor's frozen flow hypothesis) and using the velocity two-third law $D_{LL}= C_2 (\epsilon r)^{2/3}$ which is valid in the inertial range, assuming isotropy and where $C_2 = 2.0$ is a universal constant (\cite{PopeBook2000}). As the inertial range has limited extension (see Fig. \ref{fig:PSDAllAlphaWithCompensatedInsets}a), we first identify the value of $r$ such that $D_{LL}(r)/r^{2/3}$ becomes locally flat versus $r$, and then employ the velocity two-third law. We then obtain the Kolmogorov length scale $\eta = (\nu^3/\epsilon)^{1/4} = \unit{0.22}{\milli\meter}$ and the dissipation time scale $\tau_{\eta} = (\nu/\epsilon)^{1/2} = \unit{49}{\milli\second}$. Moreover, we estimate the Taylor-microscale $\lambda_u = \sqrt{15 \nu (2k/3)/\epsilon} = \unit{4.9}{\milli\meter}$ and the Taylor-Reynolds number $Re_{\lambda_u}=(2k/3)^{1/2} \lambda_u/\nu = 130$, where $k=3u_0'/2$ is the average turbulent kinetic energy, by assuming isotropy. To estimate the velocity integral length scale $L_u$, we use $\epsilon = C_{\epsilon} u_0'^3/L_u$, where $C_{\epsilon}=0.9$ (\cite{ValentePRL2012,VassilicosAnnuRev2015}).

Secondly, for the scalar field characterisation, the temperature fluctuations are measured by a thermistor. It is well-known that passive scalar fields are not isotropic at small scales when there is a large-scale mean temperature gradient (\cite{Warhaft:2000AnnuRev}) but for the order of estimation of the quantities below, we assume isotropy. First, to estimate the scalar dissipation rate $\epsilon_{\theta}$. To do so, we again assume isotropy and use the temperature (scalar) two-third law $D_{\theta\theta}(r) = \tilde{C_{\theta}}\epsilon_{\theta}\epsilon^{-1/3} r^{2/3}$ which is valid in the inertial-convective subrange (\cite{Monin75}), where $D_{\theta\theta}$ is the second order structure function of the temperature (scalar) fluctuation and $\tilde{C_{\theta}} \approx 0.25^{-1} C_{\theta}$ (\cite{Monin75}) with $C_{\theta} \sim 0.5$ the Obukhov-Corrsin constant (\cite{MYDLARSKI:1998JFM,Warhaft:2000AnnuRev}). Again we identify the plateau of the plot $D_{\theta\theta}(r)/r^{2/3}$ versus $r$ in order to locate the position of $r$ at which we employ the temperature two-third law. Second, we estimate the scalar Taylor microscale $\lambda_{\theta} = \sqrt{6 \kappa \langle T'^2 \rangle/\epsilon_{\theta}} =\unit{2.1}{\milli\meter}$ (see e.g. \cite{YasudaJFM2020}), again assuming isotropy. We find that our turbulent thermal mixing layer, for the single phase case, has P\'eclet numbers (based on both the velocity and temperature Taylor microscales) $\text{Pe}_{\lambda_u} = u_0'\lambda_u/\kappa = 870$ and $\text{Pe}_{\lambda_{\theta}}=u_0'\lambda_{\theta}/\kappa= 520$ respectively (\cite{YasudaJFM2020}).  The traditional choice of the microscale is $\lambda_u$ (\cite{MYDLARSKI:1998JFM}) but numerical simulations have uncovered that the small-scale anisotropy of the scalar field depends on the P\'eclet number based on  $\lambda_{\theta}$ (\cite{YasudaJFM2020}). So here we give both values. For the length scale of the scalar dissipation, since $\text{Pr} = 6.5 > 1$, we use the Batchelor scale to characterise the smallest scale of the passive scalar field which is given by $\eta_{\theta} = \eta \text{Pr}^{-1/2}$ (\cite{Tennekes1972Book}). As the mean temperature profile for the single-phase turbulent mixing layer follows a self-similar error-function profile (\cite{Ma:PoF1986,Kops:PoF2000}), we fitted the profile accordingly and obtain the large-scale characteristic length $L_{\theta}$ of the single-phase thermal mixing layer from the fitting parameters (see Appendix \ref{sec:Integrallengthscale} for details). The turbulent characteristics of the single-phase turbulent mixing layer are summarised in Table \ref{tab:turbSettings}.

\begin{table}
  \begin{center}
\def~{\hphantom{0}}
  \begin{tabular}{c|c|c|c|c|c|c|c|c|c}
    $\eta$ (mm)  & $\tau_{\eta}$ (ms) & $\eta_{\theta}$ (mm) & $\lambda_u$ (mm) & $\lambda_{\theta}$ (mm)& $Re_{\lambda_u}$ & $Pe_{\lambda_u}$ & $Pe_{\lambda_{\theta}}$ & $L_u$ (m) & $L_{\theta}$ (m) \\[1pt]
    $0.22$   & 49 & $0.085$ & $4.9$ &  $2.1$ & 130 & 870 & 520 & 0.04 & 0.15  
   \\
  \end{tabular}
  \caption{Relevant turbulent flow parameters of the single-phase turbulent thermal mixing layer in the current study ($\alpha = 0\%$, mean liquid flow speed = 0.5 m/s). Note that isotropy is assumed for the estimation of the above quantities except $L_{\theta}$. For the definitions of the parameters and the discussion on the isotropy, see Sec. \ref{sec:ch3ExpSetupAndMethod}.}
  \label{tab:turbSettings}
  \end{center}
\end{table}

\subsection{Measurement conditions: Bubble properties}

Bubbles are created by injecting regular air through needles, see Fig.~\ref{fig:setup}a. Using high-speed imaging in a backlight configuration we characterize the size of our bubbles for a variety of volume fractions, see the example photos Figs.~\ref{fig:setup}(b,c,d). We analyze the data by making use of the circular Hough transform to detect the individual (potentially overlapping) bubbles. We track the bubbles over time such as to obtain the mean bubble rise velocity, see Table \ref{tab:bubbleProperties_fitResults}. In Fig.~\ref{fig:setup}(e) we show the PDF of the bubble diameter for several $\alpha$ in the form of distribution charts. The mean bubble diameter obtained from the image analysis is approximately constant but there are changes in the bubble distribution over $\alpha$. 

First, we discuss the change of the bubble distribution for different $\alpha$ and the possible contributions that lead to such change. For $\alpha = 0.3\%$, when measuring the bubble diameter through image analysis, there are unresolved bubbles with sizes smaller than \unit{1.4}{\milli\meter}, thus overestimating the mean bubble diameter for that small $\alpha = 0.3\%$. The larger spread of the bubble size distribution for $\alpha =0.3\%$ may be accounted as follow. In order to produce bubbles in our setup, the array of needles at the bottom of the setup is fed by pressurised air from only one side (see Fig. \ref{fig:setup}a and also \cite{Gvozdic:2019RSI}). The case for $\alpha = 0.3\%$ is close to the minimum achievable gas volume fraction at which the pressure that pushes the air out from the array of needles may become more uneven at different needle positions (the needles that are closer to the inlet valve have larger pressure) than for the higher $\alpha$ cases. Thus, the bubbles come out with more diverse sizes than for the cases with higher $\alpha$. Moreover, higher $\alpha$ also means that there are more bubble-bubble interactions which include more coalescence of bubbles, thus contributing to the smaller spread of the bubble sizes than what happens at lower $\alpha$ cases. For $\alpha >0.3\%$, the mean bubble diameter is essentially unchanged, having the value of $d\approx\unit{2.5}{\milli\meter}$, see also Table \ref{tab:bubbleProperties_fitResults} column $d$. However, even for $\alpha > 0.3 \%$, the standard deviation of the bubble sizes changes slightly, in particular dropping from \unit{0.6}{\milli\meter} to \unit{0.4}{\milli\meter} when $\alpha$ increases from $2.2\%$ to $3.0\%$. The distributions remain to have similar shape for higher $\alpha$ except having a slight increase of width at the highest $\alpha = 4.7\%$ (see Table 2). However, this slight change of the width of the distribution for $\alpha > 0.3\%$ can be considered minor if we see the width of the bubble distribution is around \unit{0.5}{\milli\meter} for $\alpha > 0.3\%$.

For the slight decrease of the width of the bubble size distribution when $\alpha$ increases further from $0.3\%$, we can attribute to (1) the possibility that more even pressure distribution over the different needles that yields more evenly distributed sizes of bubbles when the bubbles comes out, and (2) more bubble-bubble interaction and coalescence of bubbles, which makes the spread of the sizes lower. There are also (3) the effect of incidence turbulence and (4) the effect of the cutting of the bubbles due to active grid (discussed in the next paragraph) which can also affect the bubble size distribution.

Second, we compare the variation of the mean bubble diameter in our case with other studies and discuss a possibility that leads to our observation. While the mean bubble diameter stays roughly constant over $\alpha$, other research, which also makes use of capillary needles to inject bubbles, indicates that the bubble diameter increases with $\alpha$. This includes a homogeneous bubbly flow in quiescent liquid (\cite{Colombet:2014JFM}) and a rising bubbly flow with a background upward liquid flow which is weakly turbulent (\cite{Roig:2007JFM}). Bubble swarms with a background upward liquid flow that is strongly turbulent were investigated in \cite{Almeras:2017JFM} and \cite{Prakash:2016JFM} by using the Twente Water Tunnel (\cite{poorte2002grid}), and they find also a small increase of $d$ with $\alpha$ but the investigated range of $\alpha$ was limited ($\alpha < 1\%$). It is not clear why the bubble sizes are roughly constant in our case. The active turbulent grid in our setup has a similar design but with different aspect ratio and has a smaller overall size than the one used in the Twente Water Tunnel (\cite{Almeras:2017JFM} and \cite{Prakash:2016JFM}). It is also reported in \cite{Almeras:2017JFM} and \cite{Prakash:2016JFM} that the fragmentation of bubbles caused by the active grid leads to smaller bubble sizes for high liquid mean flow. In our case, while there is a possibility that bubble diameter may increase with $\alpha$ before the bubbles entering the active turbulent grid, possibly the rotating speed of the active turbulent grid is too high that it cuts the bigger bubbles into smaller bubbles with sizes similar to those at lower $\alpha$.

Third, we obtained the aspect ratio $\chi$ of the bubbles by fitting the bubbles with ellipses in the high-speed footage. The bubble diameter analysis has a resolution limit of \unit{1.4}{\milli\meter}, while the aspect-ratio analysis uses all sizes (though the tiny bubbles of $\mathcal{O}(1)$ pixels are still not selected for analysis). Bubbles with smaller sizes, in particular for $\alpha = 0.3\%$, are more spherical bubbles because of the stronger effect of surface tension. For each $\alpha$, we use $\mathcal{O}(200)$ bubbles for fitting to obtain the aspect ratios statistics. Fig. \ref{fig:setup}f shows $\chi$ versus $\alpha$. The trend shows the mean be to almost monotonically increasing with $\alpha$, saturated at high $\alpha$. When $\alpha$ increases from $0.3\%$ to $1.0\%$, there is a considerable increase of aspect ratio $\chi$ from 1.4 to 1.7, see Table \ref{tab:bubbleProperties_fitResults}. Increasing $\alpha$ further, the aspect ratio saturates. In attempt to explain the monotonic increase of $\chi$ versus $\alpha$, one may check the Weber number and bubble Reynolds number, which are the dimensionless parameters for a freely rising single bubble. Although there is a slight increase of $d$ for $\alpha$ being from $0.3\%$ to $0.6\%$, the relative rise velocity of the bubbles decreased dramatically when $\alpha$ increases. Therefore, the bubble Reynolds number drops dramatically, see Table 2. Moreover, a lower Weber number based on the relative rise velocity $\text{We}_{V_r}$ is also resulted, see Table 2. However, the Weber number based the liquid velocity fluctuations $\text{We}_{U'}$ increases because the liquid velocity agitation dramatically increases with $\alpha$. There are also higher (1) bubble-bubble interactions as $\alpha$ increases and we have to take into account of (2) incident turbulence in our case and its interaction with the bubbles. Then, apart from the inhomogeneity of the pressure distribution at the needles that produces gas bubbles for very low $\alpha$, these two contributions may also affect the value of $\chi$ and subject to further systematic investigation. We note that in the case of freely rising bubble swarm simulation (see Fig. 5.12 in \cite{Roghair2012PhdThesis}), the bubbles with the same diameter become more spherical when $\alpha$ increases, as opposed to what we observed here. This simulation result may limit to the importance of other possible effects such as incidence turbulence as described above. Finally, we also point out that the change of shape is linked to $\alpha$ and we can not exclude that the change of aspect ratio is more important than the change of $\alpha$ itself. We will use these bubbles to provide additional stirring to the liquid.

For the cases with bubbles, since it is highly anisotropic between vertical and azimuthal directions, we estimate the velocity fluctuations of the liquid as $U' \equiv \sqrt{2\langle u_x'^2 \rangle + \langle u_z'^2 \rangle}$, where we assume $\langle u_x'^2 \rangle = \langle u_y'^2 \rangle$. We first explore the parameter space by increasing $\alpha$ from $0.0\%$ to $5.2\%$ for a fixed flow velocity of \unit{0.5}{\meter\per\second} and with our active grid turned on. Finally, the bubble Reynolds number $\text{Re}_{\text{bub}}=V_r d/\nu$ as obtained from visualization is in the range between  480 ($\alpha=0.6\%$) and 220 ($\alpha=4.7\%$). The above properties of the bubbles are summarised in Table \ref{tab:bubbleProperties_fitResults}.

\section {The energy and scalar spectra}
\begin{figure}
  \centerline{\includegraphics[width=1\columnwidth]{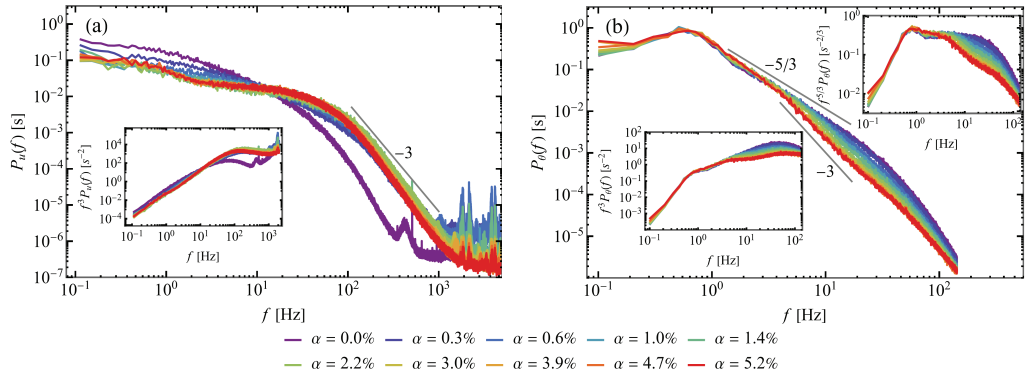}}
  \caption{ (color) $P_i(f)$, the power spectra normalised by their respective variance of (a) the velocity fluctuations ($P_u(f)$) and of (b) the temperature fluctuations ($P_{\theta}(f)$) for $\alpha$ from 0 \% (purple) to 5.2\% (red). The compensated spectra $f^3P_{i}(f)$ and $f^{5/3}P_{\theta}(f)$ are shown as insets. The scalar spectra are cut at the limit of 143 Hz, corresponding to the highest resolved frequency of the thermistor.}
\label{fig:PSDAllAlphaWithCompensatedInsets}
\end{figure}
As the scalar in the present study is advected passively by the liquid velocity, the corresponding energy spectra first need to be examined in order to understand the passive scalar spectra. In Fig.~\ref{fig:PSDAllAlphaWithCompensatedInsets} we show the energy and scalar power spectra $P_u$ and $P_\theta$ for different gas concentrations $\alpha$ normalized by their corresponding variance, corresponding to the normalized energy and scalar power spectra, respectively. The methods of calculating the spectra of the velocity and temperature fluctuations for two-phase flows are discussed in Appendix \ref{sec:SpectralMethod}. The energy spectra for single-phase shows a limited inertial range. However, whenever $\alpha > 0$, a pronounced $-3$ scaling is observed with the presence of an energy `bump' just before this $-3$ scaling when $\alpha$ is large enough, which is consistent with the results of \cite{Almeras:2017JFM}. We find that such $-3$ scaling occurs roughly between \unit{100}{\hertz} to \unit{1000}{\hertz} and has a higher energy content than the single-phase in the same frequency range. The physical mechanism of such small-scale enhanced velocity fluctuations is attributed to bubble-induced agitations, in which the velocity fluctuations produced by the bubbles are directly dissipated by viscosity (\cite{Lance:2006ht,Prakash:2016JFM,RissoAnnuRev2018}). 

Next, we examine the passive scalar (thermal) spectra. A scaling of $-5/3$ for \unit{1}{\hertz} $\leq f \leq$ \unit{15}{\hertz} is observed for the single-phase scalar spectrum. The phenomenon that a developed inertial range is observed in a passive scalar spectrum but not in the corresponding energy spectrum due to limited Reynolds number was found previously by \cite{Jayesh1994PoF}. As opposed to the relative high frequency (small-scale) fluctuation enhancement in the energy spectra, the scalar spectra show relative high frequency (small-scale) diminution when $\alpha$ increases from zero. The $-5/3$ scaling is followed by a steeper slope when $\alpha$ increases but such change of scaling is gradual and approximately saturated to a rough scaling of $-3$ for high enough $\alpha$ ($\alpha \approx 4.7 \%$) around $f=\mathcal{O}(10\text{Hz})$. However, as seen also in the inset of Fig. \ref{fig:PSDAllAlphaWithCompensatedInsets}b for $-5/3$ compensated spectra or the spectral local exponent in Fig. \ref{fig:BatchelorFitAndfcfdComparison}b, the original $-5/3$ scaling becomes less pronounced and has a steeper slope also at the frequency range before the developed $-3$ scaling for the thermal spectra. The main differences between the energy and the passive scalar spectra are that the rising bubbles produce additional fluctuations in the former but diminish the fluctuations in the latter. As we will explain below, this can be understood due to the added small-scale mixing of the thermal fluctuations by the bubbles which smoothens the temperature field. The emergence of the $-3$ scaling is abrupt in the former but gradual in the latter; and the $-3$ scaling occurs at $\mathcal{O}(100\text{Hz})$ for the former but $\mathcal{O}(10\text{Hz})$ for the latter. Clearly, there are differences in the physical mechanisms that lead to the same $-3$ scaling for the energy and passive scalar spectra. We first clarify the physical and mathematical background before speculating on the emergence of the $-3$ scaling in the scalar spectra.

The main commonalities and differences of the physical settings of the velocity and temperature fields are as follows. One common aspect is that along the measurement section, both the viscosity and diffusivity reduce the velocity and temperature fluctuations, respectively. On the other hand, the mean velocity field is homogeneous in the bulk of the measurement section while the temperature field has a large-scale mean temperature gradient. The high-$\text{Re}$ bubbles have viscous boundary layers at their air-water interface but the thermal boundary layer is expected to be minimal as the heat capacity of the bubbles is tiny, and their thermal response is fast with at most a time scale of (conductive case, no thermal convection) $t=\delta^2/D_\text{air}=\mathcal{O}(\unit{50}{\milli\second})$, where $\delta = 0.4d$ is the penetration depth for a spherical bubble and $D_\text{air}$ the thermal diffusivity of air. If convective flow inside the bubble is included this typical time-scale is drastically reduced, implying that the bubble can adjust its temperature very quickly to its surroundings. Moreover, if we compare the heat capacities of the water and the bubble we get: $(95\% C_{p,\text{water}}\rho_{\text{water}})/(5\% C_{p,\text{air}}\rho_{\text{air}}) \approx 65000$, meaning that the bubbles can not be very effective carriers of thermal energy as compared to the water. The viscous boundary layers transport momentum to the liquid phase at the scale related to bubbles, producing velocity agitations. The absence of thermal boundary layers implies no heat transport is present at the scale related to bubbles. Therefore, the only source of scalar fluctuations is from the mean temperature gradient while the source of velocity fluctuations is from bubble agitations.

Mathematically, for the spectral behavior of a passive scalar, one can consider the general spectral equation for the three dimensional scalar spectra  $E_{\theta}(k,t)$ derived from the advection-diffusion equation (\cite{Monin75}) which neglects viscous heating,
\begin{equation}\label{eq:spectralEq}
    \frac{\partial}{\partial t} E_{\theta}(k,t) + 2\kappa k^2 E_{\theta}(k,t) = T_{\theta}(k,t) + \Pi_{\theta}(k,t),
\end{equation}
where $t$ is the time and $T_{\theta}$ and $\Pi_{\theta}$ are the local net transfer and the production at wave-number $k$ respectively. One can employ Taylor's frozen-flow hypothesis to transform from the frequency domain to the wavenumber domain. In the present case, for increasing $\alpha$, the single-phase $-5/3$ scaling of the spectrum gradually transitions to a $-3$ scaling at higher $k$ (or $f$).

With the above discussed physical settings and scaling behaviors, we now speculate on the physical mechanisms on the $-3$ scaling of the scalar spectra in bubbly flows. The $-5/3$ range is considered to be the inertial range where there is negligible net local transfer, negligible production and neglected diffusivity (and viscosity) effect (\cite{PopeBook2000}). The scalar production due to the mean temperature gradient occurs at a smaller frequency than the $-5/3$ frequency range (lower than \unit{1}{\hertz}). Since the production due to the presence of bubbles is negligible as discussed previously, we speculate that $\Pi_{\theta} \approx 0$ at the scalar $-3$ subrange. Furthermore, we speculate that the rising bubbles enhance the mixing of temperature by homogenizing the temperature with increasing $\alpha$ at the scale related to the bubbles, and thus we observe a faster drop (steeper slope) in the passive scalar spectra. When \textit{$\alpha$ is large enough}, the smoothing of the temperature inhomogeneity due to the bubbles is so strong that the thermal fluctuations are \textit{directly smoothed out} by the molecular thermal diffusivity which causes the saturation of the scaling. In other words, the thermal fluctuations that originally have cascading behaviour (as in the single-phase case) passing them from larger scales to smaller scales and ceasing by the molecular thermal diffusion. When this situation happens, in this regime, we speculate that $T_{\theta} = T_{\theta} (\epsilon_{\theta},k)$, where $\epsilon_{\theta}$ is the scalar fluctuation dissipation rate. From dimensional analysis, we find that $T_{\theta} \propto \epsilon_{\theta} k^{-1}$. For a statistically stationary state, we can rearrange the terms in Eq.\ \ref{eq:spectralEq} to obtain a $E_{\theta} \propto k^{-3}$ scaling. This derivation is similar to \cite{Lance:2006ht} though they neglected the net local transfer term and performed dimensional analysis on the velocity fluctuation production by the bubbles instead. Finally, as noted above, the original $-5/3$ plateau found in Fig.\ \ref{fig:PSDAllAlphaWithCompensatedInsets}b becomes only a peak for increasing $\alpha$, i.e.\ a pure $-5/3$ scaling disappears. For high $\alpha$, even when $-3$ scaling has developed, the mixing of the bubbles may lead to an overall more homogeneous temperature field which leads to a smaller temperature fluctuation at the scales also before the $-3$ scaling, leading to a steeper slope also before that. As we will also see in the next section, the onset frequency of $-3$ scaling becomes lower for higher $\alpha$. If $\alpha$ further increases, this leads us to expect that such steeper slope (steeper than the originally $-5/3$) before the developed $-3$ scaling will be further steepened, closer to $-3$.

\section{Transition frequencies from $-5/3$ to $-3$ scaling}\label{sec:transition}

\begin{figure}
 	\centering
 	\includegraphics[width=135mm]{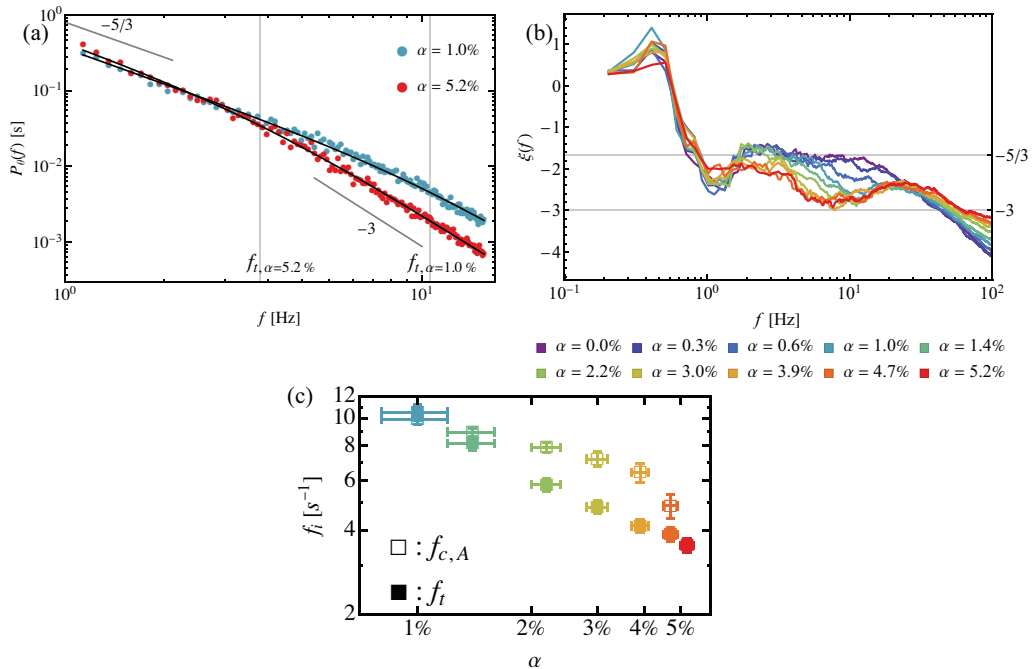} 
 	\caption{\label{fig:BatchelorFitAndfcfdComparison} (a) The power spectral densities $P_{\theta}(f)$ (normalized by their respective variance) of the temperature fluctuation from the measurements (dots) and the corresponding fits (Eq.~\ref{eq:BatchelorFit}) in logarithmic scale for the gas volume fractions $\alpha = 1.0\%$ and $\alpha = 5.2\%$ at the frequency range of \unit{1.2}{\hertz} to \unit{15}{\hertz} (colors same as Fig.~\ref{fig:PSDAllAlphaWithCompensatedInsets}), plotted together to the purpose of showing the quality of the fits. The two grey sloped lines labeled with `$-5/3$' and `$-3$' indicate the scaling behaviours of the two limits. The vertical grey lines labeled by $f_t$ indicate the fitted transition frequency for the two cases. (b) The local scaling exponent $\xi(f)$ (defined by Eq. \ref{eq:localslope}) of $P_{\theta}(f)$ over an interval of more than three decades for different gas volume fractions $\alpha$. Two grey horizontal lines indicates the the slopes $-5/3$ and $-3$. The details of obtaining the local slope is described in Section \ref{sec:transition}. (c) Comparison of $f_{c,A}(\alpha)$ (open symbols, Eq.~\ref{eq:fc}) and $f_t(\alpha)$ (solid symbols). The uncertainty of $\alpha$ is related to the precision of the differential pressure gauge (\cite{Gvozdic:2019RSI}). while the error bars of $f_d$ and $f_{c,A}$ are the confidence intervals from the fitting Eq. \ref{eq:BatchelorFit} for a $95\%$ confidence level. The the fitting results of $f_t$ are also tabulated in Table~\ref{tab:bubbleProperties_fitResults}.}
 \end{figure}

The frequencies corresponding to the onset of the $-3$ scaling of the scalar spectra are now examined. When $\alpha$ increases from 0\% to 5.2\%, the $-5/3$ scaling becomes less pronounced and, when $\alpha$ is large enough, it is followed by a $-3$ scaling, see Fig. \ref{fig:BatchelorFitAndfcfdComparison}a for examples. We identify the onset frequency $f_t$ by using the parameterization
\begin{align}\label{eq:BatchelorFit}
   P_{\theta}(f)=\frac{(f/f_L)^{\zeta_{b}}}{\left[1+(f/f_t)^2\right]^{\frac{\zeta_{b}-\zeta_{a}}{2}}}
\end{align}
which captures a transition from one scaling to another. Here $f_L$ is a fitting parameter which reflects the height of the curve, and $f_t$ is the transition frequency from the scaling $f^{\zeta_b}$ (for $f \ll f_t$) to $f^{\zeta_a}$ (for $f \gg f_t$) with scaling exponents $\zeta_b$ and $\zeta_a$, respectively, see Fig.~\ref{fig:BatchelorFitAndfcfdComparison}. Here we do not fit these exponents, but set $\zeta_b = -5/3$ (Kolmogorov-Obukhov value) (\cite{Monin75}) and $\zeta_a = -3$, consistent with the exponent of the velocity spectra in high-Re bubbly flows (\cite{RissoAnnuRev2018}) and reflecting the observed limiting cases. To determine the range that covers most of the $-5/3$ and $-3$ scaling, we examine the local logarithmic slope $\xi(f)$ of the scalar power spectra, which is given by
\begin{align}\label{eq:localslope}
    \xi(f) = \frac{d \log_{10} P_{\theta}(f) }{d \log_{10} f}.
\end{align}

To estimate the local logarithmic slope at different frequency, for each spectrum, a moving fit of a straight line in logarithmic space (centred at a frequency with a window size of one-fourth decade of the frequency) is employed in order to obtain a local logarithmic slope at that frequency. The results of $\xi (f)$ for different $\alpha$ are shown in Fig. \ref{fig:BatchelorFitAndfcfdComparison}b. In the figure, we see that around \unit{1.2}{\hertz}, $-5/3$ scaling starts to develop and the $-3$ scaling ends around \unit{15}{\hertz} for $\alpha = 2.2\%$. Therefore, \unit{1.2}{\hertz} to \unit{15}{\hertz} is a reasonable range that covers the transition from $-5/3$ to $-3$ scaling for all cases.

\begin{landscape}
\begin{table}
  \begin{center}
  \begin{tabular}{c|c|c|c|c|c|c|c|c|c|c|c}
    \hline
      $\alpha$ \unit{}{(\%)} & $U'$ \unit{}{(\meter\per\second)} & $d$ \unit{}{(\milli\meter)} & $\text{We}_{U'}$ & $\chi$ & $V_{bub}$\unit{}{(\meter\per\second)} & $\langle V_r \rangle$\unit{}{(\meter\per\second)} & $\langle \text{We}_{V_r} \rangle$ & $ \langle \text{Re}_\text{bub} \rangle$ & $f_t$\unit{}{(\hertz)} & $f_L$\unit{}{(\hertz)} & $f_{c,A}$\unit{}{(\hertz)}  \\[3pt]
      \hline
$0.3 \pm 0.2$ & --- & $2.4 \pm 0.5$* & --- & $1.4\pm 0.5$ & $0.68 \pm 0.14$ & --- & --- & --- & --- & --- & ---\\
$0.6 \pm 0.2$ & $0.042 \pm 0.008$ & $2.5 \pm 0.5$\phantom{*} & $0.06 \pm 0.03$ & $1.6 \pm 0.6$ & $0.70 \pm 0.13$ & $0.194 \pm 0.03$ & $1.30 \pm 0.04$ & $512 \pm 9$ & --- & --- & $11.0 \pm 0.4$ \\
$1.0 \pm 0.2$ & $0.051 \pm 0.007$ & $2.5 \pm 0.5$\phantom{*} & $0.09 \pm 0.03$ & $1.7 \pm 0.6$ & $0.68 \pm 0.11$ & $0.179 \pm 0.004$ & $1.13 \pm 0.04$ & $480 \pm 10$ & $10.6 \pm 0.7$ & $0.54 \pm 0.01$  & $9.9 \pm 0.4$\\
$1.4 \pm 0.2$ & $0.047 \pm 0.006$ & $2.6 \pm 0.6$\phantom{*} & $0.08 \pm 0.03$ & $1.8 \pm 0.7$ & $0.67 \pm 0.11$ & $0.164 \pm 0.003$ & $0.96 \pm 0.03$ & $448 \pm 7$ & $8.1 \pm 0.5$ & $0.55 \pm 0.01$ & $9.0\pm 0.3$\\
$2.2 \pm 0.2$ & $0.064 \pm 0.005$ & $2.6 \pm 0.6$\phantom{*} & $0.15 \pm 0.04$ & $1.9 \pm 0.7$ & $0.64 \pm 0.11$ & $0.148 \pm 0.003$ & $0.79 \pm 0.04$ & $411 \pm 9$ & $5.8 \pm 0.3$ & $0.58 \pm 0.02$ & $7.9 \pm 0.4$\\
$3.0 \pm 0.2$ & $0.073 \pm 0.006$ & $2.5 \pm 0.4$\phantom{*} & $0.19 \pm 0.04$ & $1.9 \pm 0.7$ & $0.62 \pm 0.11$ & $0.130 \pm 0.004$ & $0.59 \pm 0.04$ & $350 \pm 10$ & $4.8 \pm 0.3$ & $0.57 \pm 0.02$ & $7.2 \pm 0.4$\\
$3.9 \pm 0.2$ & $0.090 \pm 0.006$ & $2.5 \pm 0.4$\phantom{*} & $0.29 \pm 0.06$ & $1.9 \pm 0.7$ & $0.60 \pm 0.11$ & $0.117 \pm 0.005$ & $0.48 \pm 0.04$ & $320 \pm 10$ & $4.1 \pm 0.2$ & $0.60 \pm 0.02$ & $6.4 \pm 0.5$\\
$4.7 \pm 0.2$ & --- & $2.6 \pm 0.5$\phantom{*} & --- & $2.0 \pm 0.6$ & $0.59 \pm 0.11$ & $0.090 \pm 0.004$ & $0.29 \pm 0.028$ & $250 \pm 10$ & $3.9 \pm 0.2$ & $0.59 \pm 0.02$ & $4.9 \pm 0.5$\\
$5.2 \pm 0.2$  & & &  --- & --- & --- & --- & --- & --- & $3.5 \pm 0.2$ & $0.61 \pm 0.02$ & --- \\
\hline
  \end{tabular}
  \caption{Experimental parameters as a function of $\alpha$: velocity fluctuations of the liquid $U' \equiv \sqrt{2 u_x'^2 + u_z'^2 }$ with primed velocities referring to the standard deviation of these velocity components, area-equivalent bubble diameter $d$, the Weber number based on the turbulent velocity $U'$ for the bubbles $\text{We}_{U'}=\rho U'^2 d/\gamma$ (with $\gamma$ the surface tension), the aspect ratio of the bubble diameter $\chi$, the bubble rise velocity $V_{bub}$, the bubble relative (to the liquid phase) rise velocity $V_r \equiv V_{\text{bub}} - \langle u_z \rangle$, the mean Weber number based on the bubble relative rise velocity $\text{We}_{V_r}=\rho V_r^2 d/\gamma$,  the bubble Reynolds number $\text{Re}_\text{bub} = V_r d/\nu$, the fitting parameters $f_t$ and $f_L$ (below Eq.~\ref{eq:BatchelorFit}), and $f_{c,A}$ (Eq.~\ref{eq:fc}) (\cite{Almeras:2017JFM}). The bubble sizes and velocity are obtained by image analysis (see Sec. \ref{sec:ch3ExpSetupAndMethod}). *Note that for $\alpha = 0.3\%$, we slightly overestimate the mean bubble diameter and underestimate the width of the bubble size distribution because of the resolution limit of the image analysis for the bubble diameter. The aspect ratio of the bubbles are obtained by fitting the bubbles with ellipses. The uncertainty of $\alpha$ stems from the precision of the differential pressure gauge (\cite{Gvozdic:2019RSI}). For $U'$, $d$, $\text{We}_{U'}$, $\chi$, $V_{bub}$, we show the mean values and the corresponding standard deviations for the distributions. For $V_r$, $ \text{We}_{V_r} $ and $ \text{Re}_\text{bub}$, we show the corresponding mean values and the standard errors of the mean. For $f_L$ and $f_t$, we show the mean values and the corresponding confidence intervals for a 95\% confidence level obtained from the fitting. For $f_{c,A}$ , we show the mean values and the corresponding confidence intervals for a 95\% confidence level.}
  \label{tab:bubbleProperties_fitResults}
  \end{center}
\end{table}
\end{landscape}

We fit the spectra of Fig.~\ref{fig:PSDAllAlphaWithCompensatedInsets}b with Eq.~\ref{eq:BatchelorFit} with $\zeta_b=-5/3$ and $\zeta_a = -3$ only for the $\alpha \geq 1.0 \%$ cases, since for $\alpha < 1\%$ the transition frequencies resulting from the fits are either outside the fitting range or very close to the boundary of the fitting range. The fits for $\alpha = 1.0 \%$ and $\alpha = 5.2\%$ are shown in Fig. \ref{fig:BatchelorFitAndfcfdComparison}a. It shows qualitatively nice fits for $\alpha = 1.0\%$, even though this case does not exhibit a fully developed $-3$ scaling. The figure also shows that for the case of $\alpha = 5.2\%$ the scaling is close to $-3$ after the transition and that the transition frequency $f_t$ is smaller for this higher value of $\alpha$.

Table \ref{tab:bubbleProperties_fitResults} shows the transition frequencies $f_t$ obtained from the parameterisation (Eq. \ref{eq:BatchelorFit}) as function of $\alpha$ for $\alpha \geq 1.0\%$. A decreasing trend of $f_t$ with increasing $\alpha$ is found. Currently, there is no prediction for the onset frequencies $f_t$ of the bubble-induced subrange for the passive scalar spectra. However, there is a candidate of the onset frequency of the $-3$ scaling in the \textit{energy spectra}, which is proportional to $V_r/d_{eq,V}$, where $V_r$ is the relative (to the liquid phase) rise velocity of the bubbles and $d_{eq,V}$ is the mean volume equivalent diameter of the bubbles (\cite{Almeras:2017JFM}). This frequency is given by $f_{\text{bub}} \equiv V_r/\lambda_{\text{bub}} \equiv C_D V_r/d_{eq,V}$ (\cite{Almeras:2017JFM}), where $C_D$ is the drag coefficient of the bubbles, and $\lambda_{\text{bub}} \equiv d_{eq,V}/ C_D$ is the candidate for the onset scale for the $-3$ scaling in the energy spectra in the wavenumber space (\cite{RIBOUX:2009JFM,RissoAnnuRev2018}). Apart from $f_{\text{bub}}$, there are several other frequency scales that are proportional to $V_r/d_{eq,V}$, which roughly locate a `bump' in the energy spectra just before the $-3$ scaling for high enough $\alpha$, summarised in \cite{Almeras:2017JFM}. For example, based on their observations in their energy spectra, \cite{Almeras:2017JFM} gives
\begin{align}\label{eq:fc}
    f_{c,A} = 0.14 \langle V_r \rangle /\langle d \rangle,
\end{align}
where we stress again that $d$ is the \textit{area-equivalent} diameter whereas in \cite{RIBOUX:2013gd} the mean \textit{volume-equivalent} diameter is used instead.
Table \ref{tab:bubbleProperties_fitResults} includes these values of $f_{c,A}$, which we are also shown in Fig.\ \makeblue{\ref{fig:BatchelorFitAndfcfdComparison}}c. Since the $-3$ scaling has not developed for low $\alpha$ and thus the identification of $f_t$ for low $\alpha$ is hard to determine, we can only conclude that both $f_c$ and $f_t$ identified here have decreasing trends but of the same order of magnitude in our parameter regime. 

Apart from $f_{c,A}$, when $\alpha$ becomes large, the typical time scale for the effective diffusion for passive scalar in bubble-induced turbulence is the time scale for successive bubble passages (\cite{Almeras:2015JFM}), which is given by
    \begin{align}
	T_\text{2B} = \frac{d}{\alpha V_r \chi^{2/3}}.
\end{align}
The numeral values of the quantities on the right hand side of the above equation can be found in Table \ref{tab:bubbleProperties_fitResults}. For $\alpha=4.7\%$, $T_\text{2B} = \unit{0.39}{\second}$ which results in a frequency of $\unit{2.6 \pm 0.7}{\hertz}$, which is two times smaller than $f_{c,A} = \unit{4.9 \pm 0.5}{\hertz}$. The transition frequency $f_t = \unit{3.9 \pm 0.2}{\hertz}$ for the corresponding $\alpha$. For low void fraction ($\alpha=0.6\%$), we find $T_\text{2B} = \unit{1.57}{\second}$, giving a frequency of $\unit{0.7 \pm 0.3}{\hertz}$. Therefore, for high $\alpha$, this is comparable to our $f_t$. However, we see that the trend of this prediction is going up with increasing $\alpha$, while we find that the transition frequency coming from the spectra goes down with $\alpha$ (see Fig. \ref{fig:BatchelorFitAndfcfdComparison}c). This is consistent with that for large $\alpha$ the time scale for effective diffusion by bubble-induced turbulence can be dominated by the bubble passage time; while for small $\alpha$ such effective diffusive time scale can be, for example, the Lagrangian integral time scale $T_L$ (\cite{Almeras:2015JFM,Almeras2019IJMF}). The cross-over of the relevant time scale can occur when $T_{\text{2B}}= T_L$ (\cite{Almeras2019IJMF}). However, the Lagrangian integral time scales for $\alpha >0$ is not available in our study. We cannot exclude the time scales mentioned above to be relevant to the onset of the $-3$ scaling in the thermal spectra in bubbly flows.

\section{Conclusion}\label{sec:conclusion}

The above speculation on the $-3$ scaling mechanism for a scalar spectrum requires diffusivity to be important. This means that the inverse of scalar diffusive time scale needs to be around the onset frequency $f_t$ of the $-3$ subrange. A direct verification of this hypothesis may be accomplished by investigation of the diffusive and spectral transfer budgets of the scalar fluctuations at the $-3$ subrange by direct numerical simulations similar to what \cite{PandeyJFM2020} did. Our hypothesis will be confirmed if within the --3 subrange, the spectral transfer of the scalar fluctuation (behaving as $T_{\theta} \propto k^{-1}$) is balanced by the diffusive dissipation ($2\kappa k^2E_{\theta}(k)$).

 To conclude, when a swarm of high-Reynolds number bubbles are injected into a turbulent thermal mixing layer ($0.3\% \leq \alpha \leq 5.2 \%$), the scalar spectra that originally have a $-5/3$ scaling, for large frequencies develop a steeper $-3$ slope for larger gas concentration $\alpha$. As opposed to the energy spectra, which abruptly developed a $-3$ scaling with the enhancement of the small-scale content for non-zero $\alpha$, the change of spectral scaling for passive scalar spectra is more gradual by smoothing out the small-scale content when $\alpha$ increases. While the $-3$ scaling at $\mathcal{O}(100\text{Hz})$ of the energy spectra is attributed to the balance of the production of velocity fluctuations due to bubbles and the molecular viscous dissipation, we speculate that the physical mechanism of the steeper spectral slope in the passive scalar spectra is due to enhanced mixing of bubbles that promotes the homogenization of small-scale temperature differences. The scaling saturates to $-3$ around $\mathcal{O}(\unit{10}{\hertz})$ when the smoothing of temperature fluctuations is so strong such that the local net transfer of the spectral fluctuation is directly diffused by molecular diffusivity. The transition frequency $f_t$ from the $-5/3$ to $-3$ scaling in the scalar spectra are obtained from the parameterization (Eq. \ref{eq:BatchelorFit}) and found to monotonically decrease with increasing $\alpha$. 
 
 In our experiments we had $\text{Pr} = 6.5$ at \unit{22.7}{\celsius}. This Pr could experimentally be changed in a certain narrow regime by changing the water temperature (e.g. $\text{Pr} = 12.9$ and $\text{Pr} = 2.2$). To explore the Pr-dependence, direct numerical simulations would be most welcome. Note however that the much smaller Pr numbers of liquid metal (e.g. $\text{Pr} \approx 0.006$ for liquid sodium at $\unit{570}{\kelvin}$, see \cite{CioniPRE1997}) are again experimentally accessible and that the fluctuations of the temperature in turbulent bubbly sodium may be relevant for applications in working facilities in nuclear power station.\\
 
{\small

\noindent \textbf{Acknowledgments.} We thank Dennis P.M. van Gils, Gert-Wim Bruggert, and Martin Bos for the technical support. We thank Timothy Chan and Jack Cheung for preliminary measurements.\\

\noindent \textbf{Funding statement.} This work was supported by The Netherlands Center for Multiscale Catalytic Energy Conversion (MCEC), an NWO Gravitation Programme funded by the Ministry of Education, Culture and Science of the government of The Netherlands. Chao Sun acknowledges the financial support from Natural Science Foundation of China under Grant nos. 11988102 and 91852202.\\

\noindent \textbf{Declare of interests.} The authors report no conflict of interest.
}

\appendix

\section[Calculating spectra]{Method of calculating the energy and scalar spectra in two-phase flow}\label{sec:SpectralMethod}

The hot-film voltage $E_{hf}(t)$ is measured over time. We remove parts of the signal where bubbles interact with the sensor using a threshold method on the time derivative of the measured voltage across the hot-film (see \cite{RENSEN:2005hc} and the references therein) or the time derivative of the liquid velocity (\cite{Almeras:2017JFM}). During a bubble collision with a probe, there are rapid changes of $E_{hf}$, and the typical shapes of $E_{hf}(t)$ when bubbles colliding with a hot-film can be found in, for example, \cite{RENSEN:2005hc} and \cite{RensenIJMF2005partI}. In this work, by inspection of the derivative $dE_{hf}/dt$ over time, a fixed threshold $|dE_{hf}/dt| \geq \unit{500}{\volt\per\second}$ is used to detect the moments when bubbles are colliding the probe. We find that the findings and the conclusions of this work are robust against the threshold values within a reasonable range. Apart from the said threshold, we also set the minimal time between two successive bubble collisions (\unit{5}{\milli\second}) in order to determine the moments when bubbles impinging and leaving the probe and to capture the entire bubble interaction events. We check the latter threshold by examining the distributions of the bubble residence time and any unphysical bumps present in the energy spectra due to many unphysically short bubble collisions. To calculate the velocity power spectra, we use the Bartlett Method (\cite{OppenheimBook}) with linear interpolations between the `gaps' of the liquid phase velocity over time for each $\alpha$. Each partition segment (\unit{20}{\second}) cover at least one turnover time by inspecting the auto-correlation function of the signal. Linear interpolation between the gaps of the liquid signal to calculate the power spectra was also used in \cite{Almeras2016TimeResolved,Almeras:2017JFM}. The discussion on the effects of using linear interpolation can be found in \cite{RENSEN:2005hc} and \cite{Almeras2016TimeResolved}. For the optical method that calculated the energy spectra while also having evenly sampled time of the liquid velocity field, recently \cite{Ma2022JFM} employed high-resolution particle shadow velocimetry. The advantage of this method is that the two-dimensional velocity field is captured simultaneously without employing Taylor hypothesis (\cite{Ma2022JFM}). However, the method in this work only measured single-point velocity field and requires Taylor hypothesis when the time coordinate is converted into spatial coordinate (and it is only one-dimensional).

For the computation of the scalar spectra, we did not create gaps of the temperature time series to distinguish between the gas and liquid phases. This is mainly due to that the response time of the thermistor is comparable to the bubble residence time and the bubbles have similar temperature as the surrounding liquid, which is discussed and shown in \cite{DungThesis}. The Bartlett method (\cite{OppenheimBook}) is applied to the temperature signal using partition segments of \unit{10}{\sec} which cover at least one turnover time which was found by inspecting the auto-correlation function of the signal.

\section[Integral length scale]{Obtaining the thermal integral length scale}\label{sec:Integrallengthscale}

\begin{figure}
  \centerline{\includegraphics[width=70mm]{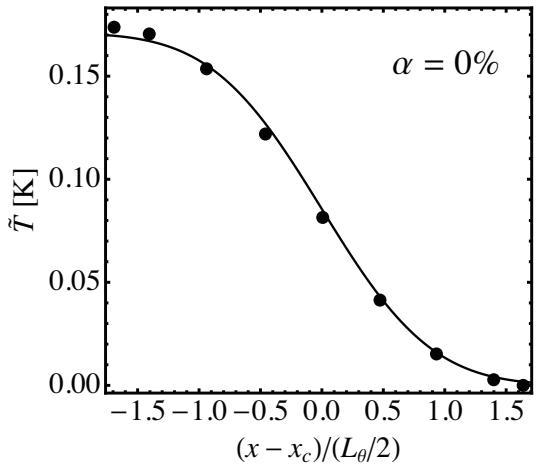}}
  \caption{The mean temperature profile $\Tilde{T}$ versus $\frac{x-x_c}{L_{\theta}/2}$ for $\alpha = 0\%$ and the corresponding fit using Eq. (\ref{eq:errorFunctFit}) (solid black line), where $\Tilde{T} \equiv  \langle T - T_{\text{inlet}} \rangle -\langle T - T_{\text{inlet}} \rangle|_{x/L = 0.9} $, $T_{\text{inlet}}$ is the temperature measured of the incoming liquid before being heated up by the heaters, $x_c = L/2$ is the x-coordinate of the middle of the tunnel, and $L_{\theta}$ is the fitting parameter that measures the width of the thermal mixing layer (Eq. (\ref{eq:errorFunctFit})).}
\label{fig:DTfitQuality}
\end{figure}

For the thermal integral length scale, intuitively it is of the order of the size of the thermal mixing layer. For a single-phase thermal-mixing layer, the mean temperature profile follows a self-similar shape that can be characterised by a complementary error function (\cite{Ma:PoF1986,Kops:PoF2000}). Therefore, quantitatively we first fit the mean temperature profile (with an offset of the mean temperature at the coldest position) $ \Tilde{T} (x)$ by the function 
\begin{align}\label{eq:errorFunctFit}
    \Tilde{T}(x)= \frac{\Delta T}{2}\text{erfc}\left(\frac{x-x_c}{L_{
\theta}/2}\right),
\end{align} 
where $x_c$ is the $x$-coordinate at the center of the measurement section, $\text{erfc}(x)$ is the complementary error function; and $\Delta T$ and $L_{\theta}/2$ are the fitting parameters which describe the asymptotic value of the mean profile and the characteristic length of the thermal mixing layer, respectively, see Fig.\ \ref{fig:DTfitQuality} for the fit. $L_{\theta}$ is then defined as the integral length scale of the scalar fluctuations. The results are summarised in Table \ref{tab:turbSettings}.

\bibliographystyle{jfm}
\bibliography{bibliography_all}

\end{document}